\documentclass[prl,aps,amsmath,amssymb,footinbib,showpacs,twocolumn,superscriptaddress]{revtex4-1}
\usepackage{graphicx}
\usepackage{dcolumn,color}
 \usepackage{multirow}
\usepackage{float}

\newcommand{\bee}{\begin{eqnarray}}
\newcommand{\ee}{\end{eqnarray}}
\newcommand{\bma}{\begin{pmatrix}}
\newcommand{\ema}{\end{pmatrix}}
\newcommand{\balig}{\begin{align}}
\newcommand{\ealig}{\end{align}}

\newcommand{\bZ}{\mathbb{Z}}
\newcommand{\ba}{\begin{align}}
\newcommand{\ea}{\end{align}}

\newcommand{\ignore}[1]{}

\usepackage{dcolumn}
\newcolumntype{C}[1]{>{\centering\let\newline\\\arraybackslash\hspace{0pt}}m{#1}}

\usepackage{subfigure}
\usepackage{bookmark}

\begin{document}
\title{Helical Majorana edge mode in a superconducting antiferromagnetic quantum spin Hall insulator}
\author{Yingyi Huang}
\affiliation{
Condensed Matter Theory Center and Joint Quantum Institute and Station Q Maryland, Department of Physics, University of Maryland, College Park, MD 20742, USA}
\affiliation{
State Key Laboratory of Optoelectronic Materials and Technologies, School of Physics, Sun Yat-sen University, Guangzhou 510275, China}
\author{Ching-Kai Chiu}
\affiliation{
Condensed Matter Theory Center and Joint Quantum Institute and Station Q Maryland, Department of Physics, University of Maryland, College Park, MD 20742, USA}
\affiliation{
Kavli Institute for Theoretical Sciences, University of Chinese Academy of Sciences, Beijing 100190, China}

\begin{abstract}
A two-dimensional time-reversal symmetric topological superconductor is a fully gapped system possessing a helical Majorana mode on the edges. This helical Majorana edge mode (HMEM), which is a Kramer's pair of two chiral Majorana edge modes in the opposite propagating directions, is robust under time-reversal symmetry protection. We propose a feasible setup and accessible measurement to provide the preliminary step of the HMEM realization by studying superconducting antiferromagnetic
quantum spin Hall insulators. Since this antiferromagnetic topological insulator hosts a helical electron edge mode and preserves effective time-reversal symmetry, which is the combination of time-reversal symmetry and crystalline symmetry, the proximity effect of the conventional $s$-wave superconducting pairing can directly induce a single HMEM. We further show the HMEM leads to the observation of an $e^2/h$ conductance, and this quantized conductance survives even in the presence of small symmetry-breaking disorders.

\end{abstract}
\date{\rm\today}
\maketitle

 The discovery of the quantum spin Hall (QSH) effect~\cite{PhysRevLett.95.226801,Bernevig1757,Molenkamp,Taylortheory} has boosted the study of and search for topological materials and opened the door to fundamentally new physical phenomena and their potential applications for novel devices. The QSH insulator, also known as a time-reversal symmetric topological insulator, characterized by $\bZ_2$ invariant~\cite{Kane:2005kx,Fu2007uq}, hosts a helical electron mode localized on the edges of the insulator~\cite{Hart:2014aa}.
 This helical mode is stable under time-reversal symmetry and leads to quantized two-terminal longitudinal charge conductance $G=2 e^2/h$~\cite{Molenkamp,Z2Shinsei,chiu_RMP_16,Qi2008sf}, which is robust against disorder scattering and other perturbation effects. This concept of robust boundary modes protected by symmetries stimulates the development of another branch of the topological materials -- topological superconductors (TSC).

There has been an enormous interdisciplinary interest in TSCs in the community. The first generation of the TSC is a one-dimensional (1D) superconductor (SC) hosting Majorana bound states on the wire ends~\cite{Sau_semiconductor_heterostructures,Roman_SC_semi,Gil_Majorana_wire}. A Majorana bound state, which is its own antiparticle, has zero energy locked by particle-hole symmetry. The observation of the zero-bias-peak conductance has provided tentative evidence supporting the Majorana bound state existence~\cite{Mourik_zero_bias,RokhinsonLiuFurdyna12,Deng_zero_bias,Churchill_zero_bias,Das_zero_bias,Finck_zero_bias}. Recently, another important breakthrough~\cite{He294} showed the potential evidence of the second TSC generation as a 2D TSC hosting chiral Majorana edge modes through observing the plateau of $0.5e^2/h$ two-terminal conductance in the hybrid quantum anomalous Hall (QAH)-superconductor device~\cite{PhysRevB.83.100512,PhysRevB.92.064520,PhysRevB.82.184516}.

These two types of the aforementioned Majorana modes in Altland-Zirnbauber symmetry class D~\cite{altlandZirnbauerPRB10} are not the only two Majorana species. According to the classification table of topological insulators and superconductors~\cite{SchnyderAIP,Kitaev2009,chiu_RMP_16,PhysRevLett.102.187001}, the third TSC generation is realized as a 2D time-reversal symmetric TSC hosting 1D helical Majorana edge modes (HMEMs) in symmetry class DIII. An HMEM, composed of two chiral Majorana edge modes with the opposite propagating directions as a Kramer's pair, cannot hybridize in the presence of time-reversal symmetry. The possible realization of HMEMs has been proposed in unconventional superconductors with exotic superconducting pairing~\cite{PhysRevB.79.094504,PhysRevLett.108.147003,PhysRevB.90.054503,PhysRevB.94.214502,PhysRevLett.111.056402} and the interface of two $s$-wave superconductors with a $\pi$ phase difference~\cite{PhysRevB.83.220510,Teo:2010fk}; due to the subtlety and uncertainty of the superconducting pairing in unconventional superconductors~\cite{PhysRevB.89.220510,RevModPhys.75.657} and the difficulty of controlling the $\pi$-phase without breaking time-reversal symmetry, the observation of the HMEMs have not become reality.

\begin{figure}
\centering
\includegraphics[width=1.0\columnwidth]{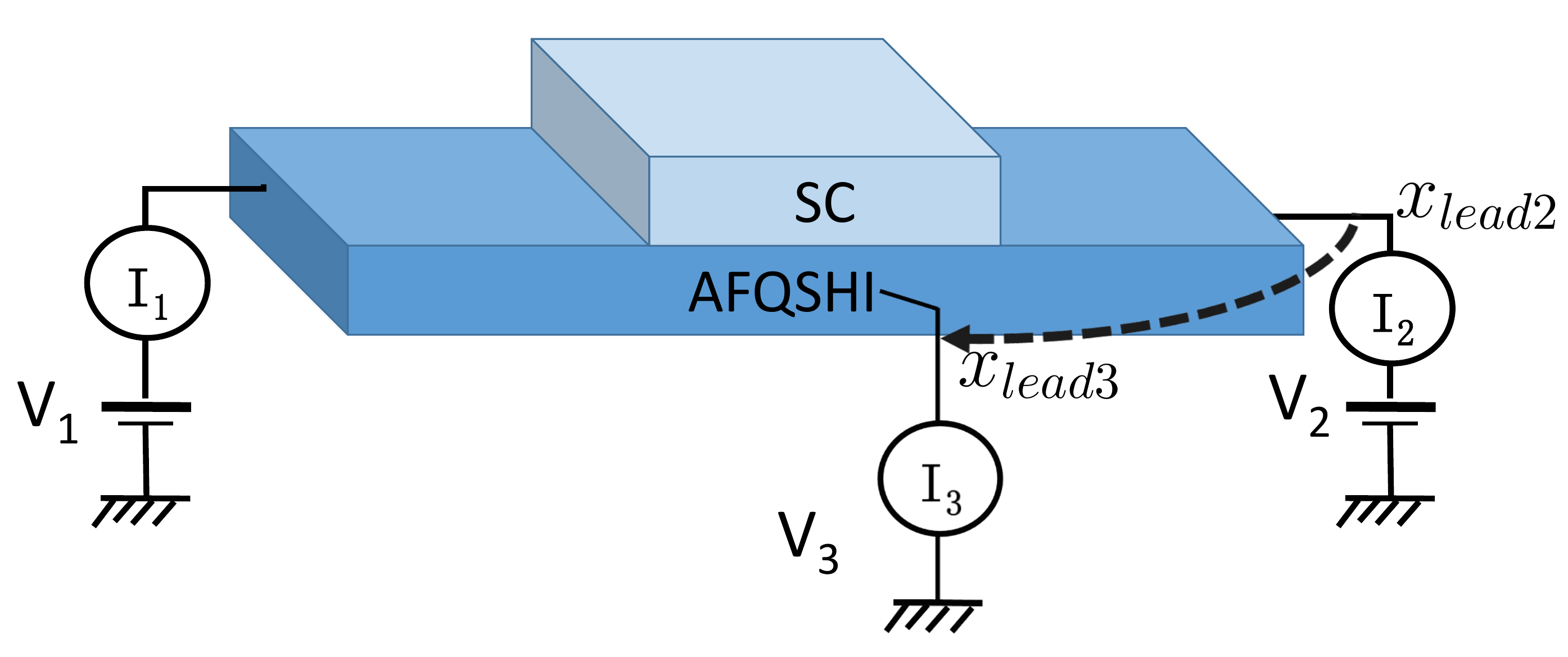}
\caption{(Color online) Schematic of the hybrid AFQSHI-SC device shows an $s$-wave SC bar deposited on the top of the AFQSHI central region. The two-terminal conductance can be measured between the two of the normal leads. Label $V_i$ denotes the voltage of lead $i$ and $V_3$ is set to zero. The arrow marked the position of the right lead, which can be between the position of lead 2 ($x_{lead2}$) and the position of lead 3 ($x_{lead3}$). }
\label{device}
\end{figure}

In this manuscript, we propose a feasible alternative experimental realization route for HMEMs by engineering a conventional $s$-wave superconductor and the recent transport measurement, which was successfully used to support the existence of chiral Majorana edge modes~\cite{He294}. The main difference between the helical and chiral Majorana setups is that the QAH film is replaced by a 2D antiferromagnetic quantum spin Hall insulator (AFQSHI)~\cite{AFTI,2017arXiv170707427N}, preserving \emph{effective} time reversal symmetry. This effective time reversal symmetry can circumvent the requirement of the exotic superconducting pairing to realize an HMEM.
We show that an HMEM is present on the edge of the $s$-wave superconducting-proximitzied AFQSHI as an antiferromagnetic topological superconductor (AFTSC). The setup we propose is, as shown schematically in Fig.~\ref{device}, the AFQSHI-SC hybrid system hosting the HMEM, which leads to $e^2/h$ conductance. Furthermore, since an AFQSHI has been experimentally supported by scanning tunneling spectroscopy measurement in a superconductor FeSe/SrTiO3(001) film~\cite{wang2016topological}, our proposal avoids the obstacles of the previous proposals and directly pave the way to realize HMEMs.

\begin{figure}[t!]
\centering
\includegraphics[width=\columnwidth]{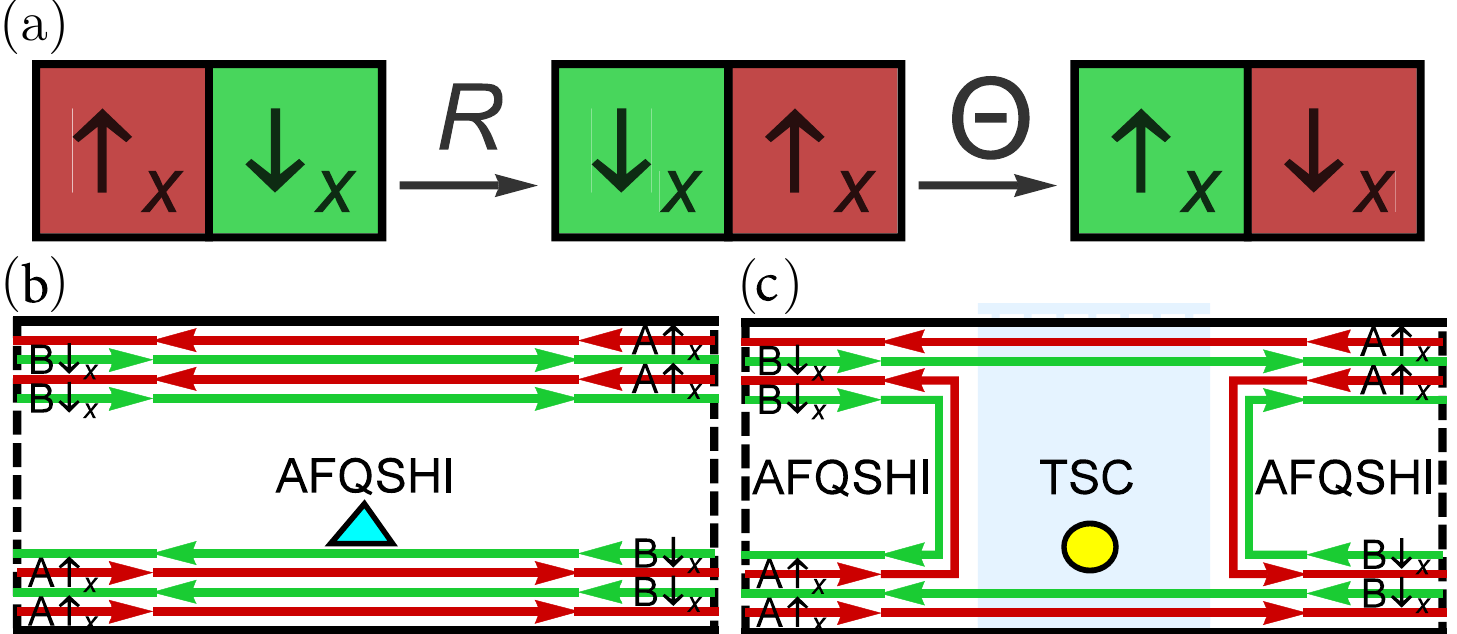}
\caption{(Color online) (a) The operation of the effective time reversal symmetry $\tilde{\Theta}=\Theta R$ on atom $A$ (in green) with spin pointing up in $x$ direction and atom $B$ (in red) with spin pointing down in $x$ direction in one unit cell. We note A and B atoms are identical particles with the opposite spin directions. While the time reversal operation $\Theta$ flips the spin, $R$ switches atom $A$ and $B$. (b) The flow diagram of the HMEMs in the AFQSHI without superconductivity. Label $x$ in the arrows indicates $1/2$-spin pointing to the $x$ direction. (c) The flow diagram of the HMEMs in the SC-AFQSHI device. Two HMEMs move together on the edges of the AFQSHI and are separated on the corners of the superconducting region (in light blue color). The cyan triangle and yellow circle indicate the trivial and the non-trivial phases as shown in Fig.~\ref{phase_diagram}(a). }
\label{AF}
\end{figure}

AFQSHI, which breaks time reversal symmetry, relies on an \emph{effective} time reversal symmetry $\tilde{\Theta}=\Theta R$~\footnote{See Sec.~I of the Supplemental Material for the conditions of the effective time reversal symmetry.}, which combines the time reversal symmetry $\Theta$ with the exchange $R$ of the two atoms in the unit cell~\cite{AFTI}, as illustrated in Fig.~\ref{AF}(a). Since $\tilde{\Theta}^2=-1$ and $\tilde{\Theta}$ acts as the effective \emph{non-spatial} symmetry operation by interchanging atoms only in the same unit cell, the system belongs to class AII and is characterized by $\bZ_2$ invariant, similar to QSH effect~\cite{Kane:2005kx,Fu2007uq}. The non-trivial phase possesses a helical electron edge mode protected by this effective time reversal symmetry $\tilde{\Theta}$.
	
We start with the Hamiltonian as the effective AFQSHI model written in form of
\bee
\hat{\mathcal{H}_0}=\sum_\textit{k} \psi_\textbf{k}^\dag\textsl{H}_0(\textbf{k})\psi_\textbf{k},
\ee
where annihilation operators are given by $\psi_\textbf{k}=(c^A_{\textbf{k}\uparrow},c^A_{\textbf{k}\downarrow},c^B_{\textbf{k}\uparrow},c^B_{\textbf{k}\downarrow})^T$ and
\begin{equation}
\textsl{H}_0(\textbf{k})=M(k)\tau_z\sigma_z+\sin k_x\tau_0\sigma_x+\sin k_y\tau_0\sigma_y.
\end{equation}
Pauli matrices  $\tau_\alpha$ and $\sigma_\beta$ represent sublattice and $1/2$-spin degrees of freedom, respectively, $M(k)\equiv m_0+\cos k_x+\cos k_y$, and the lattice constant $a=1$. This Hamiltonian, which  preserves effective time-reversal symmetry with the symmetry operator $\tilde{\Theta}=i\tau_x\sigma_y\mathcal{K}$, obeys $\tilde{\Theta}\textsl{H}_0(-\textbf{k})\tilde{\Theta}^{-1}= \textsl{H}_0(\textbf{k})$ and breaks the physical time reversal symmetry.
 On the one hand, in the trivial region ($|m_0|>2$), the AFSQHI does not possess stable gapless states on its edges. On the other hand, in the topological region ($0<|m_0|<2$), the non-zero $\bZ_2$ invariant defined by $\tilde{\Theta}$ leads to a helical electron edge mode, which is a Kramer's pair of the opposite chiral electron modes. As shown in Fig.~\ref{AF}(b-c), one of the AFQSHI essential features is that these two chiral edge modes, with the opposite spins in different atoms, cannot be directly gapped in the presence of the conventional superconducting $s$-wave pairing~\footnote{See Sec.~II of the Supplemental Material for comparison between the regular QSHI and AFQSHI in the presence of $s$-wave SC.}.
As the superconductivity proximity is induced, the survival of the helical electron mode is the essential factor for the HMEM presence.

As the $s$-wave superconductor is deposited on the top of the AFQSHI, as illustrated in Fig.~\ref{device}, due to the SC proximity effect, a finite superconducting pairing amplitude is induced in the AFQSHI. The superconducting AFQSHI model is described by the Bogoliubov-de Gennes (BdG) Hamiltonian $\mathcal{H}_\text{BdG}=\sum_\textbf{k}\Psi^\dag_\textbf{k}\textit{H}_\text{BdG}\Psi_k/2$, where in Nambu basis $\Psi_\textbf{k}=[(c^A_{\textbf{k}\uparrow},c^A_{\textbf{k}\downarrow},c^B_{\textbf{k}\uparrow},c^B_{\textbf{k}\downarrow}),(c^{A\dag}_{-\textbf{k}\uparrow},c^{A\dag}_{-\textbf{k}\downarrow},c^{B\dag}_{-\textbf{k}\uparrow},c^{B\dag}_{-\textbf{k}\downarrow})]^T$
and
\begin{equation}
\textit{H}_\text{BdG}(\textbf{k})=\begin{pmatrix}H_0(\textbf{k}) & -i\Delta \tau_0 \sigma_y \\ i\Delta\tau_0\sigma_y & -H_0^*(-\textbf{k}) \end{pmatrix}.
\label{BdGeq}
\end{equation}
 For the superconductor deposit, it is ideal to choose a conventional superconductor possessing the intra-orbital $s$-wave pairing, such as Niobium~\cite{PhysRev.139.A1515} or others. Through superconductor proximity effect, the intra-orbital $s$-wave pairing ($\Delta= \langle c^{A\dag}_{\textbf{k}\uparrow} c^{A\dag}_{-\textbf{k}\downarrow} \rangle = \langle c^{B\dag}_{\textbf{k}\uparrow} c^{B\dag}_{-\textbf{k}\downarrow} \rangle$) dominates and the inter-orbital pairing ($ \langle c^{A\dag}_{\textbf{k}\uparrow} c^{B\dag}_{-\textbf{k}\downarrow} \rangle$=$ \langle c^{B\dag}_{\textbf{k}\uparrow} c^{A\dag}_{-\textbf{k}\downarrow} \rangle$) is suppressed. This BdG Hamiltonian, which preserves particle-hole symmetry and effective time-reversal symmetry, obeys
\begin{align}
C\textit{H}_\text{BdG}(-\textbf{k})C^{-1}=&-\textit{H}_\text{BdG}(\textbf{k}), \\
T\textit{H}_\text{BdG}(-\textbf{k})T^{-1}=&\textit{H}_\text{BdG}(\textbf{k}),
\end{align}
where the particle-hole symmetry operator and effective time-reversal symmetry are given by $C=\rho_x \mathcal{K}$ and $T=\rho_0\tilde{\Theta}$, of which  $\rho_\alpha$ is the Pauli matrices represents particle-hole degrees of freedom. Thus, the 2D system belongs to class DIII and can possess non-zero $\bZ_2$ topological invariant, leading to an HMEM. Even in the presence of the weak inter-orbital pairings, which preserve the symmetries, the HMEM is not affected as long as the $\bZ_2$ invariant is non-zero. Since the real time reversal symmetry $\Theta$ and the reflection symmetry $R$ are broken, this SC system is completely different from the reflection symmetric superconductor in class DIII~\cite{Kane_Mirror,ChiuSchnyder14}.

To study the topology of the superconductor model, the BdG Hamiltonian can be rewritten in a block diagonal form
\begin{equation}
\textit{H}_\text{BdG}'(\textbf{k})=\begin{pmatrix} \textit{H}_{\rm{A}}(\textbf{k}) & 0 \\0 & \textit{H}_{\rm{B}}(\textbf{k}) \end{pmatrix}.\label{blockdiag}
\end{equation}
The sub-Hamiltonians are given by
\begin{align}
H_A(\textbf{k})&=\sigma_0 \otimes  h_+(k) -\Delta  \sigma_z \otimes \sigma_z, \\
H_B(\textbf{k})&=\sigma_0 \otimes  h_-(k) -\Delta  \sigma_z \otimes \sigma_z,
\end{align}
where $h_\pm=\sin k_x \sigma_x + \sin k_y  \sigma_y \pm M(k)  \sigma_z$ and
in the basis $\frac{1}{\sqrt{2}}(c^{\eta\dag}_{\textbf{k}\uparrow}+c^\eta_{-\textbf{k}\downarrow},c^{\eta\dag}_{\textbf{k}\downarrow}+c^\eta_{-\textbf{k}\uparrow},c^{\eta\dag}_{\textbf{k}\uparrow}-c^\eta_{-\textbf{k}\downarrow},c^{\eta\dag}_{\textbf{k}\downarrow}-c^\eta_{-\textbf{k}\uparrow})^T$ with $\eta=A, B$. This block-diagonalized property is accidental since in the presence of other symmetry-preserving terms, the BdG Hamiltonian might not be able to be written in this simple form. Because the topology of the system in class DIII is described by $\bZ_2$ invariant ($N_{\bZ_2}$), we only need to consider the Chern number of one block, say $H_{\rm{A}}$, which is the time reversal partner of $H_{\rm{B}}$, to determine trivial (even Chern number) or non-trivial (odd Chern number) topology.

\begin{figure}[t!]
\centering
\includegraphics[width=1.0\columnwidth]{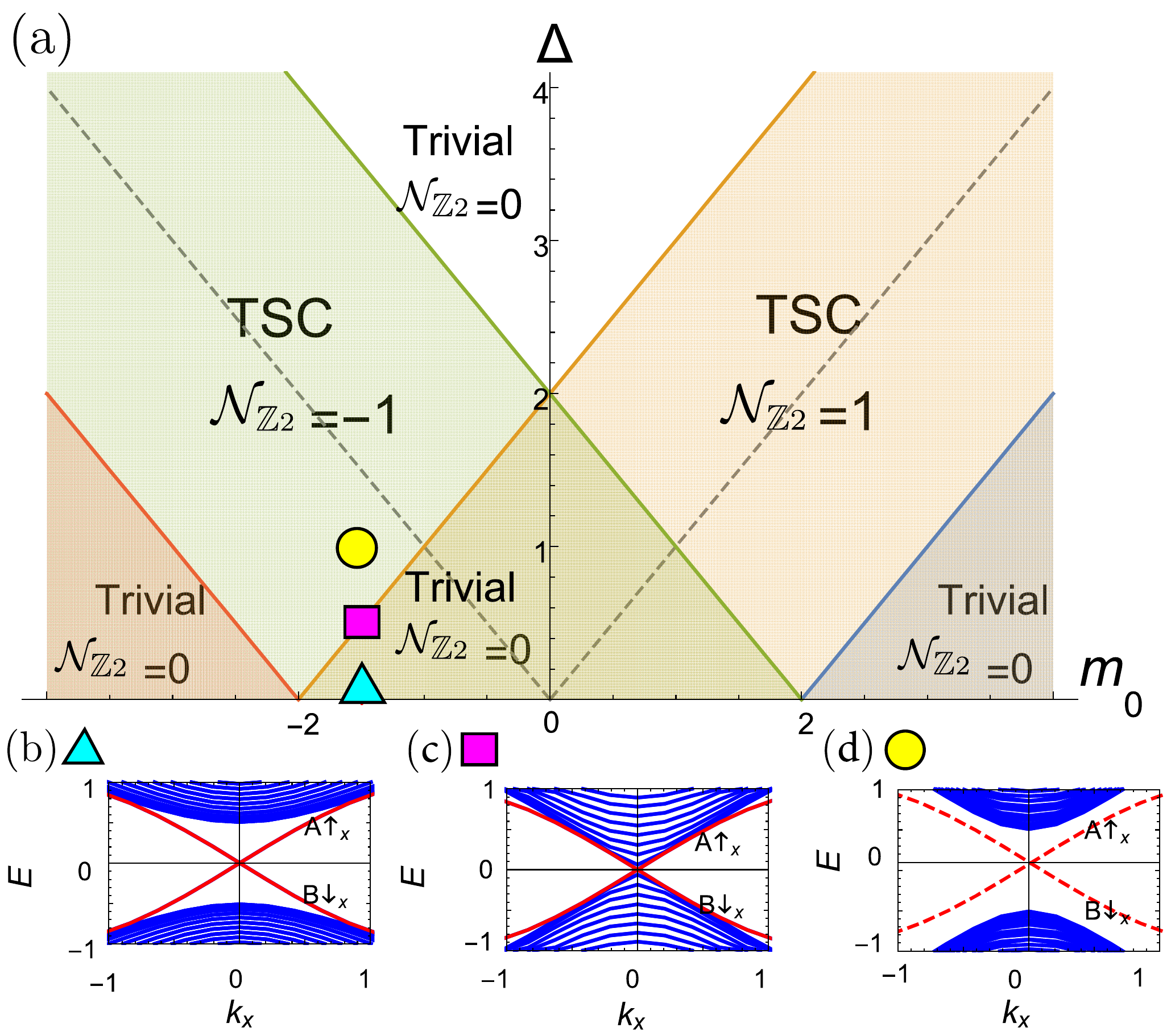}
\caption{\label{phase_diagram}(Color online) (a) Phase diagram of AFQSHI-SC system.
(b)-(d) The energy spectra with $m_0=-1.5$ and $\Delta=0.0, 0.5, 1.0$ measured at the positions labelled by cyan triangle, magenta square and yellow circle in (a), respectively. While the red solid lines in (b) and (c) indicate the helical electron (two Majorana) edge mode. At the topological phase point (c), one of the HMEMs is eliminated by the bulk gap closing; then, one HMEM  (the red dashed lines in (d)) remains.}
\end{figure}
	A point touching of the linear bulk band closing, as topological phase transition, changes the $\bZ_2$ invariant by 1. Since the square of the $H_A$ energy spectrum is given by
\begin{equation}
E^2_\pm=(M(k)\pm\Delta)^2+\sin^2k_x+\sin^2k_y,
\end{equation}
the linear band closing at one point (Fig.~\ref{phase_diagram}(c)) is located at $m_0+\Delta= \pm 2$ and $m_0-\Delta=\pm 2$ ($m_0\pm \Delta=0$ are not the topological phase transition point due to the band closing at two points, which cannot change the $\bZ_2$ invariant.) In the atomic insulator limit, the region of $\Delta=0$ (Fig.~\ref{phase_diagram}(b)) and $|m_0|\gg 1$ is obviously in the trivial phase; hence, the non-trivial phase is in the two regions: (a) $m_0+\Delta>-2,\ m_0+\Delta<2,\ m_0-\Delta<-2$, (b) $\ m_0+\Delta>2,\ m_0-\Delta>-2,\ m_0-\Delta<2$, as shown in Fig.~\ref{phase_diagram}(a)~\footnote{See Sec.~III of the Supplemental Material for alternative approach to obtain the phase diagram by calculating $Z_2$ invariant with inversion symmetry.}. In this non-trivial phase, the subsystem $H_A$ hosts a chiral Majorana edge mode, while the subsystem $H_B$ as the $H_A$ time-reversal partner hosts another chiral Majorana edge mode with the opposite propagating direction. Hence, the entire system is an AFTSC hosting a single HMEM as shown in Figs.~\ref{AF}(c) and  \ref{phase_diagram}(d) with the $\bZ_2$ invariant $N_{\bZ_2}=1$.

For an HMEM observation, the setup we propose is an $s$-wave superconductor on the top of the AFQSHI central region and the two opposite sides of the AFQSHI in contact with two leads separately, as shown in Fig.~\ref{device}. The transport measurement in this setup is accessible since in the similar setup, the plateau of $0.5e^2/h$ conductance, which is the potential evidence of a chiral electron edge mode, has been observed~\cite{He294}. We first analytically study the transport property in this setup as the superconductor in the central region possesses an HMEM ($N_{\bZ_2}=1$). Generalized Landauer-B\"uttiker formula provides the expression of the current flowing from lead $i$ at zero temperature as~\cite{Lambert,Datta}
\begin{equation}
I_i=\sum_{j=1,2} g_{ij}(V_j-V_{3})
\label{Landauer}
\end{equation}
where
\begin{equation}
g_{ij}=\frac{e^2}{h}(N_i\delta_{ij}-\mathcal{T}^N_{ij}+\mathcal{T}^A_{ij}).
\label{gij}
\end{equation}
Here, $V_j$ is the voltage on lead $j$ and $V_3$ is the voltage of the SC, which is set to be 0, and $N_i$ is the effective number of conducting channels in contact with lead $i$. Since the HMEM can split into two chiral conducting channels, we have $N_1=N_2=2$. For $i\neq j$, $\mathcal{T}^{N(A)}_{ij}$ is the normal (Andreev) transmission coefficient of an electron in lead $j$ to be transmitted to lead $i$ as an electron (hole). For $i=j$, $\mathcal{T}^{N(A)}_{ii}$ is the normal (Andreev) reflection coefficient of an electron in lead $i$ to be reflected back to itself as an electron (hole). The coefficients obey the normalization condition $\sum_j T^N_{ij}+T^A_{ij}=N_i$.

\begin{figure}
\centering
\includegraphics[width=1.0\columnwidth]{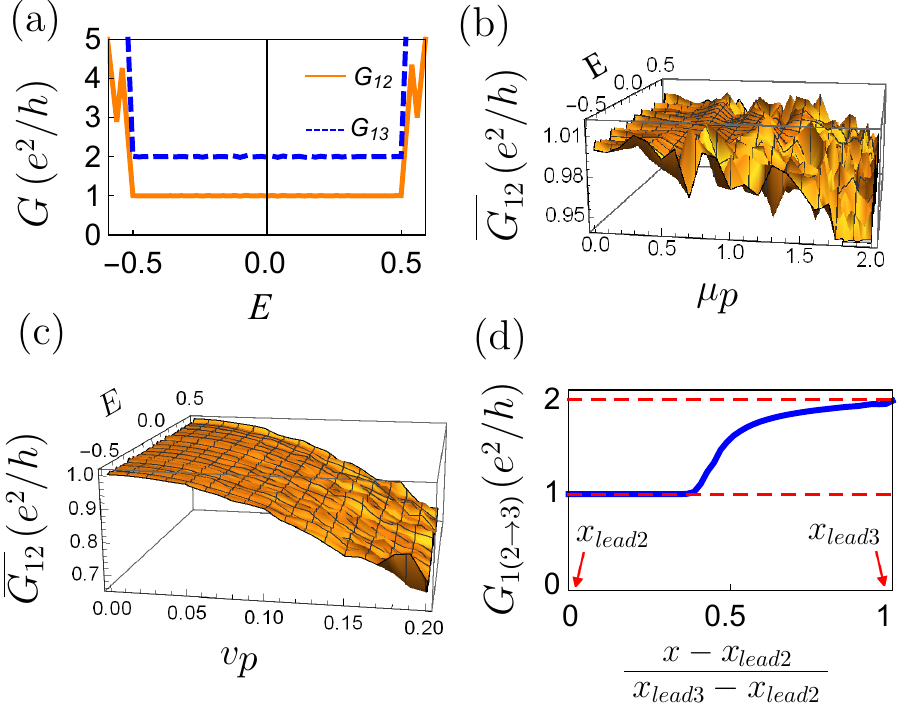}
\caption{\label{conductance} (Color online) (a) The two terminal differential conductances $G_{12}$ (solid orange line) and $G_{13}$ (dashed blue line) through the AFQSHI-SC structure as illustrated in Fig.~\ref{device}. The parameters adopted are the hybridization gap $m_0=-1.5$, the superconductoring pairing potential $\Delta=1$, the lattice size of the AFQSHI $150a\times30a$, the superconductor size in the central AFQSHI region $50a\times30a$.
(b,c) The average conductances $G_{12}$ of 500 samples in the presence of the two different disorder types.
In the region $[-\mu_p,\mu_p]$ the sublattice potentials $\mu_{\bf r}^A$ and $\mu_{\bf r}^B$ randomly distribute in the real space.
 The second type of the symmetry breaking term $v_{\bf r}\rho_z \tau_x \sigma_z$ distributes in the real space with random value $v_{\bf r}$ in the region $[-v_p, v_p]$. (d) The conductance $G_{1 (2\rightarrow3)}$ varies as a function of $x$, the position of the right lead, which is between the position $x_{lead2}$ and $x_{lead3}$ marked in Fig.~\ref{device}.
 }
\end{figure}

When no current flows to the lead 3 $(I_3=0)$, by using the current conservation $I_1=-I_2$ and Eq.~\eqref{Landauer}, the two-terminal conductance is given by
\begin{equation}
G_{12}\equiv\frac{I_1}{V_1-V_2}=\frac{g_{11}g_{22}-g_{12}g_{21}}{g_{11}+g_{22}+g_{12}+g_{21}}.
\label{G12}
\end{equation}
When the current flows through lead 3, instead of lead 2, the conditions $I_2=0$ and $I_1=-I_3$ and Eq.~\eqref{Landauer} lead to
\begin{equation}
G_{13}\equiv\frac{I_1}{V_1-V_3}=g_{11}-\frac{g_{12}g_{21}}{g_{22}}.
\label{G13}
\end{equation}
Similarly,
\bee
G_{23}\equiv\frac{I_2}{V_2-V_3}=g_{22}-\frac{g_{12}g_{21}}{g_{11}}.
\ee
The device can be considered as two hybrid QAH-SC systems, each of which hosts one chiral electron edge mode in the opposite propagating direction. Hence, the reflection and transmission coefficients of AFTSC are double the coefficient values of the hybrid QAH-SC system~\cite{PhysRevB.83.100512}: $\mathcal{T}^{N(A)}_{ij}=1/2 $ for $i,j=1,2$. It is expected that the conductance values of the hybrid AFQSHI-SC device
\bee
G_{12}=\frac{1}{2}G_{13}=\frac{1}{2}G_{23}=\frac{e^2}{h} \label{quantized}
\ee
are observed to support HMEM existence. Fig.~\ref{conductance}(a) shows from our numerical calculations that the $e^2/h$ conductances persist in the wide regions of the voltage difference between lead 1 and 2 ($2E\equiv V_1-V_2$). As $E$ reaches to the superconducting gap, since the current flows via the quasiparticles and quasiholes above the gap, the conductances are no longer quantized.

	The main difference between the original topological superconductor in class DIII and the AFTSC is that the effective symmetry of the AFTSC, protecting the HMEM, is crystalline symmetry combined with time reversal symmetry, instead of time reversal symmetry. In reality, due to the imperfection of the crystal growth, the crystalline symmetry is not always preserved. For example, the potential in atom A and B might not be identical in each unit cell in the presence of the potential disorder. Although the HMEM loses the symmetry protection, we further examine if the HMEM can survive once the symmetry is preserved on average~\cite{Fu:average_symmetry} maybe too strong. To verify the HMEM survival, we further compute the conductances in the presence of the disorders by using the scattering matrix method. Its numerical implementation is carried out through KWANT \cite{kwant}, which is a Python package for numerical transport calculations on tight-binding models Eq.~\eqref{BdGeq}. Two types of symmetry breaking disorder might be present in the device.
The first type is spin-independent sublattice potential randomly distributed in the entire hybrid system ($-\mu^A_{{\bf r}} c^{A\dag}_{\bf{r}}c^{A}_{\bf{r}},\ -\mu^B_{{\bf r}} c^{B\dag}_{\bf{r}}c^{B}_{\bf{r}}$). Since the system can be decomposed to two independent hybrid QAH-SC systems in class D, the sublattice potential disorders do not alter the values of the conductances.  As shown in Fig.~\ref{conductance}(b), the quantized conductances persist untill the disorder strength is much greater than the superconducting gap.
Another reason is that even if small sublattice potential uniformly distributes in real space, in the presence of this symmetry-breaking term, the HMEM survives.
On the other hand, we introduce a symmetry-breaking term that can directly gap the helical electron mode as the second type of the disorders.
We introduce more destructive disorders $v_{\bf r} \rho_z \tau_x \sigma_z$, since this term hybridizes the two chiral electron edge modes of the AFQSI in the absence of the superconductivity. This disorder with random value of $v_{\bf r}$ is distributed in the lattice Hamiltonian \eqref{BdGeq} in real space and its spatial average $\langle v_r\rangle$ is zero. Fig.~\ref{conductance}(c) shows that the conductance is close to be quantized~\footnote{See Sec.~IV of the Supplemental Material for further study of conductance at small disorder strength for different system lengths.} in the presence of the small disorders although the conductance decreases as the disorder strength grows.
	
	 It is known that the observation of $G_{12}=e^2/h$ conductance can stem from QAH effect, exhibiting one chiral electron edge mode, instead of the HMEM. However, the QAH can be excluded by observing $G_{13}=2e^2/h$ conductance from the lead in touch with the superconductor, while for the QAH the $e^2/h$ conductance should be observed, regardless the locations of the edges in contact with the leads. Although the observed conductance quantizations in both QAH~\cite{chang2013experimental} and superconducting QAH~\cite{He294} experiments are not expected to be as good as in an integer quantum Hall device~\cite{PhysRevLett.45.494}, another important evidence to support the HMEM is that as lead 3 moves toward the superconductor, the conductance gradually increases from $e^2/h$ to $2e^2/h$, as shown in Fig.~\ref{conductance}(d).

	We show a single HMEM can arise in the $s$-wave superconducting proximitized AFQSHI. Since choosing the AFQSHI can directly circumvent the requirement of the unconventional superconducting pairing~\cite{PhysRevB.79.094504,PhysRevLett.108.147003,PhysRevB.90.054503,PhysRevB.94.214502} or time-reversal symmetric $\pi$ phase difference~\cite{PhysRevB.83.220510,Teo:2010fk}, our proposal of the HMEM realization is more feasible and accessible. The potential drawback of the proposal is that the effective time-reversal might be broken due to the imperfection of the crystal growth; hence, we show the quantized conductances in Eq.~\ref{quantized} persists in the presence of small symmetry-breaking disorders. It is expected that with \emph{almost} clean crystal growth the quantized conductances, which are the HMEM signature, should be observed. Although the observation of the quantized conductances is not conclusive evidence to support the HMEM existence~\cite{CME,CME2} , our proposal is a preliminary step in accepting the HMEM in a reasonable setup probed by well-developed transport measurement.

The authors are indebted to Q.-L. He, C.-X. Liu, J. D. Sau, and F.~Setiawan for discussions. Y.H. acknowledges the funding from China Scholarship Council. CKC is supported by Microsoft Q and LPS-MPO-CMTC and the Strategic Priority Research Program of the Chinese Academy of Sciences, Grant No.~XDB28000000. We acknowledge the University of Maryland supercomputing resources (http://www.it.umd.edu/hpcc) made available in conducting the research reported in this paper.

Note added -- While submitting this manuscript, we became aware of \cite{PhysRevB.96.161407}, which shares the similar idea of effective time reversal symmetry.

\bibliography{TOPO3_v14}

\onecolumngrid
\newpage
\vspace{1cm}
\begin{center}
{\bf\Large Supplemental Material}
\end{center}
\vspace{0.1cm}
\setcounter{secnumdepth}{3}
\setcounter{equation}{0}
\setcounter{figure}{0}
\renewcommand{\theequation}{S-\arabic{equation}}
\renewcommand{\thefigure}{S\arabic{figure}}
\renewcommand\figurename{Supplementary Figure}
\renewcommand\tablename{Supplementary Table}
\newcommand\Scite[1]{[S\citealp{#1}]}
\makeatletter \renewcommand\@biblabel[1]{[S#1]} \makeatother
\section{The conditions of the effective time reversal symmetry}
\begin{figure}[H]
\centering
\includegraphics[width=0.72\columnwidth]{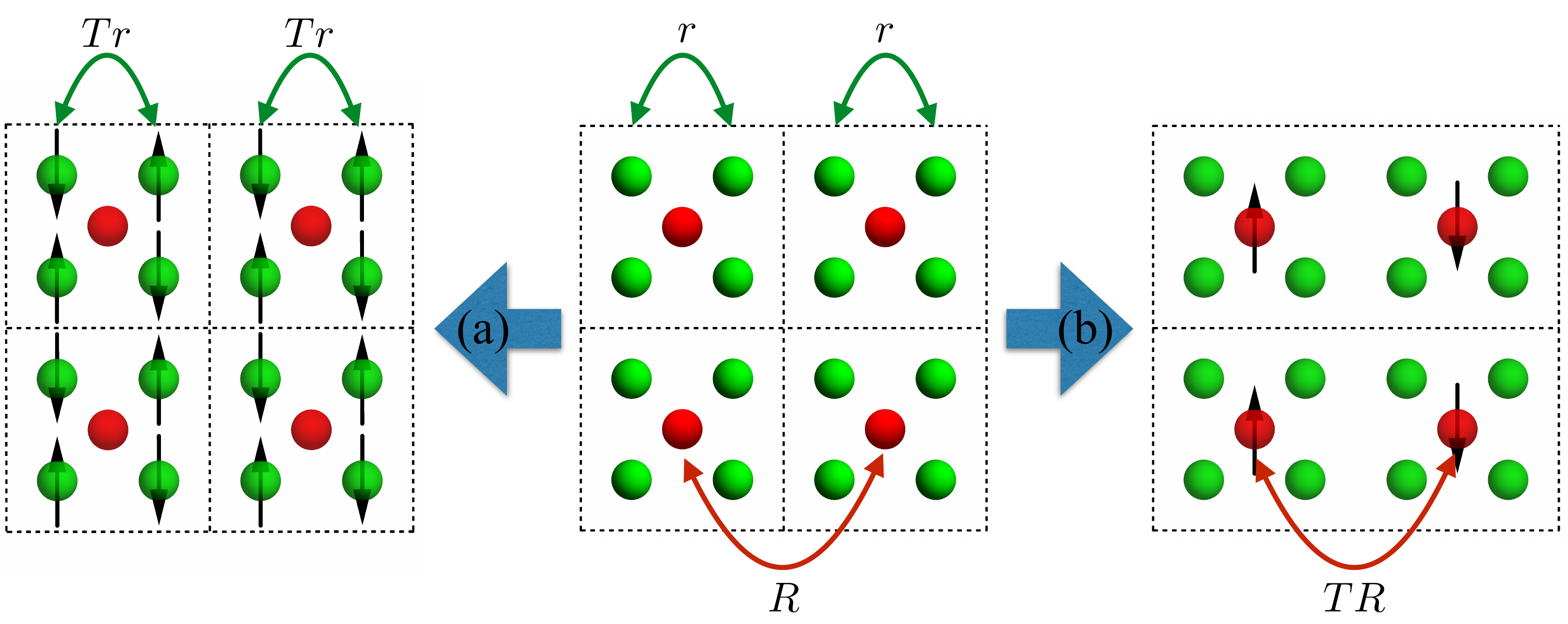}
\caption{\label{effective T} (Color online) Crystal atomic structures with and without antiferromagnetic orders. The middle panel represents a crystal atomic structure preserving physical time reversal symmetry in the absence of any antiferromagnetic order. The dashed lines indicate the boundaries of the unit cells. Four green atoms in the unit cell are identical and interchangeable indicated by the green arrow lines. The left panel represents the interchangeable green atoms possessing alternating spin order as the antiferromagnetic order, which breaks the time reversal symmetry and the interchange symmetry. Without the physical time reversal symmetry, the effective time reversal symmetry ($Tr$) is preserved. The right panel represents red atoms possessing alternating spin order along only one direction. In the presence of this antiferromagnetic order, the new unit cell includes the two original unit cells. Only the effective time reversal symmetry $(TR)$ is preserved, where $R$ represents the interchange of the red atom as red arrow lines.  }
\end{figure}

The key to realize a helical electron edge mode is the effective time reversal symmetry without the physical time reversal symmetry.
This effective symmetry can emerge in two possible arrangements of antiferromagnetic orders in crystal atomic structures. (a) in the absence of the antiferromagnetic order, the system preserves the time reversal symmetry $T H(-k) T^{-1}=H(k)$ and interchange symmetry $r H r^{-1}= H$. The interchange symmetry represents the system is invariant under interchange of the two identical atoms (say A and B) in the different locations of the same unit cell. Once the antiferromagnetic order is introduced, the time reversal symmetry is broken. To have the effective time-reversal symmetry, the antiferromagnetic order must be applied to the two interchangeable atoms with the opposition spins ($A_{i,j}^\dagger\sigma_z A_{i,j}-B_{i,j}^\dagger\sigma_z B_{i,j} $) as shown in the left panel of Fig.~\ref{effective T}. Only the effective time reversal symmetry, which is the composite symmetry $(Tr)$ of time reversal operation and the interchange of the two atoms in the same unit cell, is preserved. (b) without antiferromagnetic order, only the time reversal symmetry is required to be preserved and any two unit cells are interchangeable due to the translational symmetry. In the presence of the antiferromagnetic order, along only one direction each two closest unit cells have the opposite spins $(C^\dagger_{2j}\sigma_z C_{2j}-C^\dagger_{2i+1,j}\sigma_z C_{2i+1,j})$; the time reversal symmetry is broken and the new unit cell is extended to the two original unit cells as shown in the right panel of Fig.~\ref{effective T}. The antiferromagnetic system preserves only the effective time reversal symmetry, which is the composite symmetry $(TR)$ of time reversal operation and the interchange of the two original unit cells.  


\section{\texorpdfstring{$S$}{S}-wave superconducting quantum spin Hall insulator}
\begin{figure}[H]
\centering
\includegraphics[width=0.72\columnwidth]{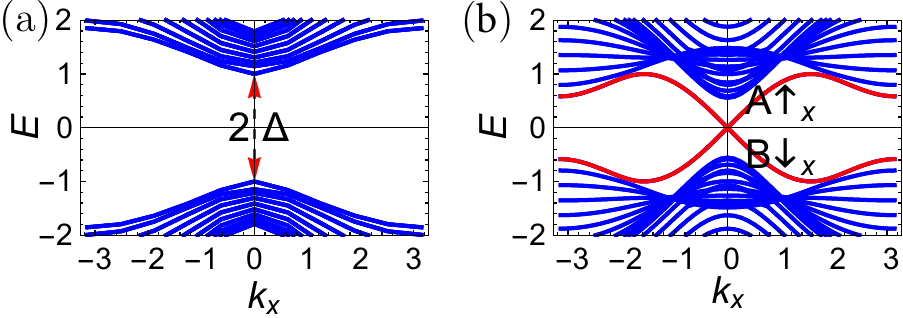}
\caption{\label{gap} (Color online) Comparison of the energy spectrum. (a) In the superconducting quantum spin Hall insulator, the helical electron edge is gapped by the conventional $s$-wave pairing. (b) In the AFTSC, the helical electron mode survives in the $s$-wave pairing since the two chiral modes are located at the two atoms (A and B) of the different locations. }
\end{figure}

Selecting AFQSHI as a platform is the key to host a HMEM in the presence of s-wave intra-orbital superconductor pairing. The reason is that although a conventional quantum spin Hall insulator possesses a helical electron edge mode protected by time reversal symmetry, the helical mode can be easily gapped in the presence of the $s$-wave intra-orbital pairing.
To show this, we consider a generic low-energy edge Hamiltonian of the quantum spin Hall insulator exhibiting linear dispersion
\bee
H_{\rm{edge}}= k_x \sigma_x \label{ferroH0},
\ee
where $\sigma_\alpha$ represent spin-$1/2$ index. That is, the two chiral edge modes are in the same orbital, which is the main difference from the AFQSHI. The system \eqref{ferroH0} preserves time reversal symmetry, obeys $\Theta\textsl{H}_{\rm{edge}}(-k_x)\Theta^{-1} =\textsl{H}_{\rm{edge}}(\textbf{k})$ with time-reversal symmetry operator $\Theta=i\sigma_y \mathcal{K}$. The time-reversal symmetry protects a helical electron edge mode, which composes of two HMEMs with the opposite spins. We include the proximity effect from the $s$-wave SC, the BdG Hamiltonian can be written as
\begin{equation}
\textit{H}_\text{BdG}(\textbf{k})=\begin{pmatrix}H_{\rm{edge}}(k_x) & -i\Delta\sigma_y \\ i\Delta\sigma_y & -H_{\rm{edge}}^*(-k_x)\end{pmatrix}.
\end{equation}
The gapless edge states are directly destroyed by the SC pairing since the edge energy spectrum $E_{\pm}=\pm \sqrt{k_x^2+ \Delta^2}$ as shown in Fig.~\ref{gap}. Therefore, the system is in the trivial phase and HMEMs are always absent. Similarly, we can consider the bulk Hamiltonian of the superconducting quantum spin Hall insulator
\begin{equation}
\textit{H}_\text{BdG}(\textbf{k})=\begin{pmatrix}H_{\rm{bulk}}(\textbf{k}) & -i\Delta\tau_0\sigma_y \\ i\Delta\tau_0\sigma_y & -H_{\rm{bulk}}^*(-\textbf{k})\end{pmatrix},
\end{equation}
where $\textsl{H}_{\rm{bulk}}(\textbf{k})=M(k)\tau_z\sigma_0+\sin k_x\tau_x\sigma_z+\sin k_y\tau_y\sigma_0$ describes the BHZ Hamiltonian~\cite{PhysRevLett.96.106802} and $\tau_\beta$ and $\sigma_\alpha$ for the orbital and spin degree of freedom respectively.
Diagonalizing the bulk BdG Hamiltonian yields the bulk energy spectrum
\begin{equation}
E_\pm =\pm \sqrt{M(k)^2+\Delta^2+\sin^2k_x+\sin^2k_y}.
\end{equation}
The bulk gap is always open in the presence of the superconducting $s$-wave pairing. Since without superconductivity QSHI is in the trivial phase in class DIII, the system is kept in the trivial phase as the pairing increases from zero. Thus, it is not possible to realize any HMEM in a superconducting quantum spin Hall insulator.

\section{\texorpdfstring{$\bZ_2$}{Z2}  invariant}
We use the bulk gap closing condition of the energy spectrum in the manuscript to identify the phase boundary of the phase diagram between the topological phase and trivial phase. Here we provide an alternative method to determine the topological number $N_{\mathbb{Z}_2}$ of the different phases, we can make use of the inversion symmetry of the system~\cite{PhysRevB.76.045302}.
The parity operator of the system is found to be $\textsl{P}=\rho_z\tau_z\sigma_z$, satisfies $H(-\textbf{k})=PH(\textbf{k})P^{-1}$. The parity eigenvalue $\xi_{2m}(\Gamma_i)=\pm1$ of $P$ of the $2m$-th occupied energy band at time-reversal invariant momentum $\Gamma_i$'s in the Brillouin zone shares the same eigenvalue $\xi_{2m}=\xi_{2m-1}$ with its Kramers degenerate partner.  In two dimensions, there are four $\Gamma_i$ points, which are at $\textbf{k}=(0,0),(0,\pi),(\pi,0)$ and $(\pi,\pi)$, respectively. They lead to
\begin{equation}
\delta_i=\prod^n_{m=1}\xi_{2m}(\Gamma_i)
\label{deltai}
\end{equation}
with $2n$ being the number of occupied energy band, whose product at special point $\Gamma_i$'s gives the $\mathbb{Z}_2$ invariants
\begin{equation}
(-1)^{\mathcal{N}_{\mathbb{Z}_2}}=\prod_i \delta_i.
\label{Z2}
\end{equation}
Table~\ref{Table:Z2} shows the calculation of $\mathbb{Z}_2$ invariants from $\delta_i$'s in different regions. Firstly, we can see that $N_{\bZ_2}$ are zero except in two regions (a) $m_0+\Delta>-2, m_0+\Delta<2, m_0-\Delta<-2$ and (b) $m_0+\Delta>2, m_0-\Delta>-2, m_0-\Delta<2$, which are topological phases. Second,  since two of the $\delta$'s change at $m_0\pm\Delta=0$, these two lines are not the boundaries between topological regions and trivial regions. This approach has the consistent  results with the approach using the bulk gap closing condition in the manuscript.

\begin{table}[H]
\centering
\begin{tabular}{ c c c c c c }
\hline
  \hline
  region &   $\delta_1$        & $\delta_2$      & $\delta_3$       & $\delta_4$      & $N_{\bZ_2}$ \\
   \hline
  $m_0+\Delta<-2$,$m_0-\Delta<-2$ & 1 & 1 & 1 & 1 & 0 \\
  $m_0+\Delta>-2$,$m_0+\Delta<0$,$m_0-\Delta<-2$ & -1 & 1 & 1 &1 & 1 \\
  $m_0+\Delta>0$,$m_0+\Delta<2$,$m_0-\Delta<-2$ & -1 & -1 & -1 & 1 & 1 \\
  $m_0+\Delta>2$,$m_0-\Delta<-2$ &  -1 & -1 & -1 & -1 & 0 \\
  $m_0+\Delta>-2$,$m_0+\Delta<0$,$m_0-\Delta>-2$,$m_0-\Delta<0$ & 1 & 1 & 1 & 1 & 0 \\
  $m_0+\Delta>0$,$m_0+\Delta<2$,$m_0-\Delta>-2$,$m_0-\Delta<0$  & 1 & -1 & -1 & 1 & 0 \\
  $m_0+\Delta>2$,$m_0-\Delta>-2$,$m_0-\Delta<0$ & 1 & -1 & -1 & -1 & 1 \\
  $m_0+\Delta>0$,$m_0+\Delta<2$,$m_0-\Delta>0$ & 1 & 1 & 1 & 1 & 0 \\
  $m_0+\Delta>2$,$m_0-\Delta>0$,$m_0-\Delta<2$ & 1 & 1 & 1 & -1 & 1 \\
  $m_0+\Delta>2$,$m_0-\Delta>2$ & 1 & 1 & 1 & 1 & 0 \\
  \hline
  \hline
\end{tabular}
\caption{\label{Table:Z2} The values of $\delta_i$ at different $\Gamma_i$(i=1,2,3,4) points and the resulting $\mathbb{Z}_2$ invariants in different regions. $\Gamma_i$ points are at momenta $\textbf{k}=(0,0),(0,\pi),(\pi,0)$ and $(\pi,\pi)$ respectively. $\delta_i$'s and $\mathbb{Z}_2$ invariants are given by Eqs.~\eqref{deltai} and \eqref{Z2} respectively. The listed ten regions are divided by $m_0\pm\Delta=\pm 2,0$.}
\end{table}

\section{The robustness of the conductance quantization}

In the main text, we have considered two different types of disorders stemming from the crystal imperfection in reality. These disorders break the effective time-reversal symmetry locally and preserve the symmetry on average. In particular, once the second type ($v_\textbf{r}\rho_z\tau_x\sigma_z$), which anticommutes with the AFQSHI Hamiltonian (Eq.~\ref{BdGeq}), becomes uniformly distributed, the HMEM is gapped. In this regard, when the strength $v_p$ of the randomly-distributed disorder reaches $0.2$, the near-zero-energy conductance $G_{12}$ for the system length $L=150a$ (in the unit of lattice constant $a=1$) is lower than the quantized value as shown in Fig.~\ref{conductance}(c). This suppression of conductance indicates that the HMEM suffers from backscattering. To realize the HMEM, it is important to study how the physics of the scattering length is affected by the disorders.

\begin{figure}
\centering
\includegraphics[width=0.9\columnwidth]{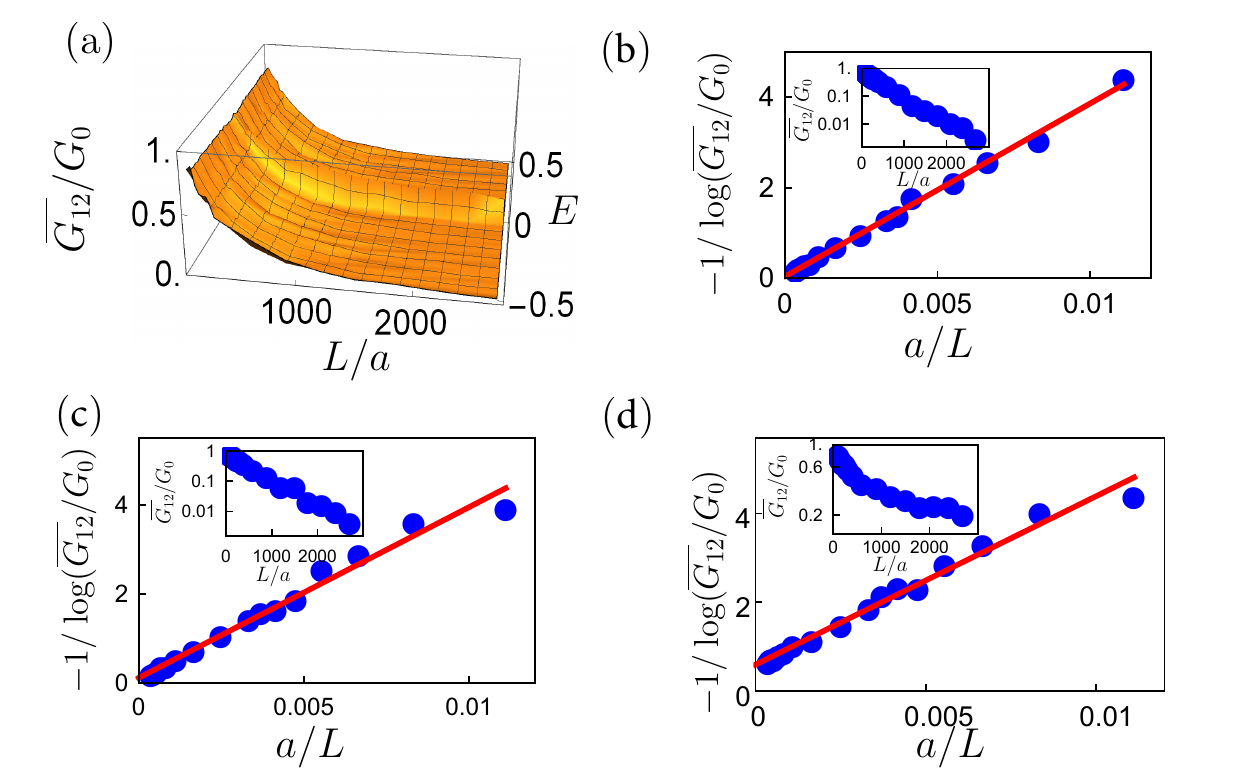}
\caption{\label{Sfig3} (Color online) The disorder-averaged two-terminal conductance $\overline{G_{12}}$ as a function of system length $L$ with fixed disorder strength $v_p=0.2$. (a) Conductances for different energy. (b-d) $-1/\log{\overline{{G_{12}}/G_0}}$ vs $1/L$ at three different energy levels: (b) $E=0.2$, (c) $E=0.02$, (d) $E=0$.  The red lines indicate linear fitting $y=\alpha+\beta x$ with coefficients for panel (b) being $\alpha=0.04(3)$, $\beta=382(6)a$, those for panel (c) being $\alpha=0.1(1)$, $\beta=383(15)a$ and those for panel (d) being $\alpha=0.6(1)$, $\beta=380(13)a$. Insets: The semi-log plot of conductance as a function of $L$. All conductances are expressed in unit of $G_0=\frac{e^2}{h}$ and all the lengths are in the unit of lattice constant $a=1$. (d) shows the conductance does not vanish as $L\rightarrow \infty$. Except for the system length, the physical parameters are identical to the ones in Fig.~\ref{conductance}.}
\end{figure}

First, we compute the two terminal differential conductances in the presence of the disorders $\rho_z\tau_x\sigma_z$ for different system lengths ($L$) with fixed disorder strength $v_p=0.2$ and average the conductance in different spatial distributions of the disorders. As shown in Fig.~\ref{Sfig3}(a), the conductance, at different energy levels within the bulk gap, decays as the system length $L$ increases. To describe the decay tendency, for small $L$, we assume that $G_{12}\propto e^{-L/\beta}$ due to the fast decay, where $\beta$ is the effective scattering length; while for large $L$, $G_{12}$ is a $L$-independent constant. That is, the conductance might be non-zero as $L\rightarrow \infty$ as shown in the inset of Fig.~\ref{Sfig3}(d). Accordingly, this decay tendency of conductance with increasing $L$ can be written in this form
\bee
G_{12}\propto e^{-\frac{1}{\alpha+\beta/L}} \label{conductance_decay}
\ee
To show the conductance behavior, at different energy levels $E=0.2, 0.02, 0.0$, we plot $-1/\log(\overline{{G_{12}}}/G_0)$ as a function of $1/L$ in Fig.~\ref{Sfig3}(b-d). Using the linear regression to fit Eq.~\ref{conductance_decay}, we estimate the value of $\beta(\sim 400a)$ determining the scattering length. Hence, in the presence of small disorder the HEME survives, as the length of experimental setup can be much shorter than the scattering length. On the other hand, comparing the fitting results of Fig.~\ref{Sfig3}(b,c) and (d), we find that the scattering lengths at the three different energy levels are almost identical for small $L$ as shown in the insets of Fig.~\ref{Sfig3}(b,c) and (d). However, for large $L$ the conductance is moved away from zero as the energy is close to zero. This can be explained by the fact that the edge modes close to zero energy, which provide electron tunnels, are protected by particle-hole symmetry.

\begin{figure}
\centering
\includegraphics[width=0.5\columnwidth]{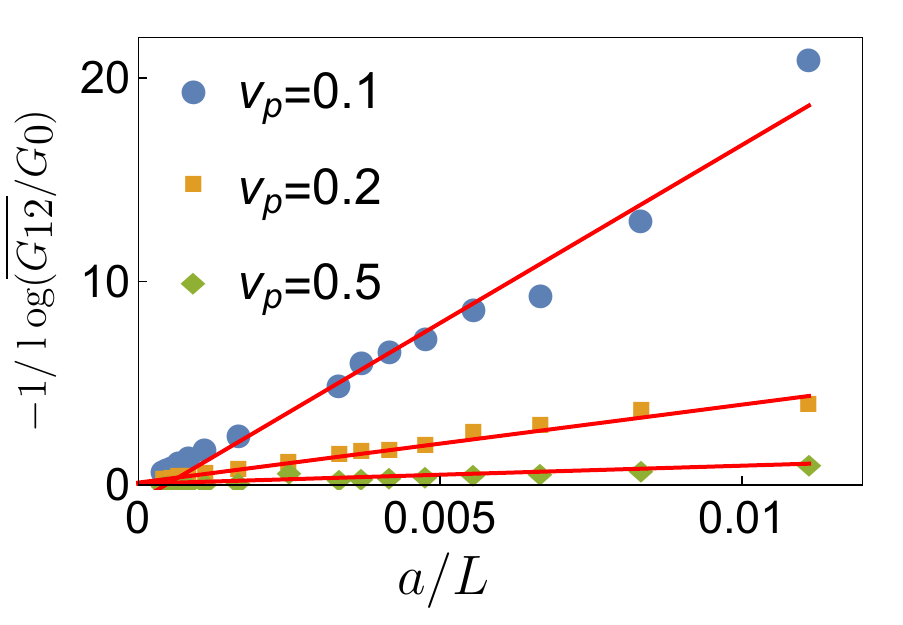}
\caption{\label{Sfig4} (Color online) $-1/\log(\overline{G_{12}}/G_0)$ at fixed energy $E=0.02$ as a function of system length $1/L$ for different disorder strength $v_p$. The red lines mark the linear regression $y=\alpha+\beta x$ with coefficients for $v_p=0.1$ being $\alpha=-0.8(6)$, $\beta=1749(126)a$, those for $v_p=0.2$ being $\alpha=0.12(7)$, $\beta=383(15)a$ and those for $v_p=0.5$ being $\alpha=0.08(7)$, $\beta=66(15)a$.}
\end{figure}

To study the relationship between scattering length and disorder strength, we then calculate the conductance at fixed energy $E=0.02$ for different disorder strengths $v_p$.
As shown in Fig.~\ref{Sfig4}, the conductance for different disorder strengths exhibit  exponential relations with the systems size. Accordingly, we can use a similar data analysis as that on Fig.~\ref{Sfig3}. We find that the estimated scattering length for $v_p=0.1,0.2,0.5$ is $1749a$, $383a$ and $66a$ respectively. That is, a stronger disorder strength leads to a shorter scattering length.

\end{document}